# The Conformal Origin of the Lepton Flavor Mixing Matrix

Cora Aigle and Alex Jourjine[1]

FG CTP
Dresden, Germany

Abstract

Using the formalism of the flavor spin theory, we re-derive the $TM_1$ lepton flavor mixing factor decomposition in a novel way, using a previously unreported near identity $4|U_{e2}U_{e3}|+|U_{e1}|^2=1$. The resulting mixing matrix is a direct consequence of the conformal symmetry of the massless SM before the EW phase transition.

In addition to the previously reported results, we derive a relation between the CP violating phase $\delta_{CP}$ and the three mixing angles, $\cos\delta_{cp}=W(\theta_{12},\theta_{13})\,\mathrm{ctg}(2\theta_{23})$. Application of the relation to the experimental data suggests the $TM_1$ decomposition should use $\delta_{cp}=(19/12)\pi$. With this assumption all four parameters of $TM_1$ may be expressed in radicals, which might help in the separation of the classical values of the mixing parameters from their quantum corrections in the framework of the flavor spin theories. This in turn could help in clarification of the mass generation mechanism.

[1] jourjine@pks.mpg.de

## 1. Introduction

Separation of the tree level and the quantum corrections for the SM parameter set is typically impossible. Coupling constants, masses and flavor mixing values cannot be in principle be derived within the SM and their experimental values seem like some random numbers. There is, however, an exception for the values of $U_{PMNS}$ in the lepton sector.

Unlike the quark sector, the patterns in the quark and lepton flavor mixing are clearly discernible and are seemingly clear [1]. For a long time the TBM approximation of $U_{PMNS}$ was considered as very good and possibly exact. The nice feature of the TBM is that its elements are algebraic. Since $U_{PMNS}$ arises as the result of spontaneous symmetry breakdown in the Higgs sector, such algebraic form suggests that its values arise from some relatively simple not yet known additional algebraic terms in the SM Lagrangian. Discovering such terms could help with understanding of the related problems, such as the hierarchy problem. This hope was reinforced with the discovery that angles $\theta_{12}, \theta_{13}$ have an algebraic expression in radicals [2]. A specific form of the lepton mixing martrix, $TM_1$, was proposed in [3] as a good data fit, after the results of the Daya Bay experiment were published. A form of $TM_1$ that arises in the flavor spin theory generically was proposes in [4], using purely theoretical considerations. Among many others [5], the relationship between the angles $\theta_{12}, \theta_{13}$ was later realized in a SM Lagrangian extension, using a discrete flavor symmetry group [6].

In this Letter, after re-deriving the known results in a novel way, by using a previously unreported near identity among the elements of the first row of $U_{PMNS}$, we derive an equation for $\delta_{cp}$ in terms of the three mixing angles. We then apply it to the experimental data sets and discuss how the results can help to determine the neutrino mass ordering. We do not address here the issue of how to generate the thus fixed matrix elements. This issue will be addressed elsewhere. Here we will only mention that flavor spin theories are severely restricted in the fermionic field content. The restriction rules out the flavor spin versions of the constructions based on discrete symmetry groups involving sizable additional fermionic content [5, 6, 7]. Essentially, the simplest flavor spin theory has just one additional Dirac field that can be used to realize the values of the lepton mixing matrix that are not generic. This field must decouple from the observable spectrum after the SSB in the Higgs sector.

In following Section 2, we re-derive $TM_1$ from the first principles, obtaining the already reported results for $\theta_{12}, \theta_{13}$ and the equation for $\delta_{CP}$. We show that all mixing angles and $\delta_{CP}$ can be expressed in radicals, provided we take $\delta_{cp} = (19/12)\pi$. The following Section 3 contains a summary and a discussion.

## 2. The Conformal Realization of the Lepton Flavor Mixing

The conformal form of $TM_1$ was derived in [4] as unexpected result of representing the fermionic degrees of freedom in the SM as quantum differential forms within the framework of the flavor spin theories. In such theories the classical mass matrix terms are restricted to the elements of $SU(2,2)$ [4, 8 - 10]. This group determines the structure of flavor mixing both for quarks and leptons. $SU(2,2)$ is isomorphic to the conformal group $SO(2,4)$, the symmetry group of the massless SM. As a result, we can speak of the conformal representation of the flavor mixing. This representation results in two tree level constraints on mixing matrices in the quark sector and one in the lepton sector. In the lepton sector the constraints is $U_{\mu 1} = U_{\tau 1}$. The constraints are not generated by discrete symmetry groups or additional horizontal gauge fields, although a horizontal gauge field with conformal gauge group and flat connection arises naturally.

It turns out that for leptons the tree level mixing can be described by $SU(1,2) \subset SU(2,2)$ [9]. For example, the general TBM matrix is given by

$$U_{TBM} = \begin{bmatrix} 1 & 0 & 0 \\ 0 & \cos\chi_{23} & -\sin\chi_{23} \\ 0 & \sin\chi_{23} & \cos\chi_{23} \end{bmatrix} \begin{bmatrix} x & y & 0 \\ z & w & 0 \\ 0 & 0 & 1 \end{bmatrix} = \begin{bmatrix} x & y & 0 \\ -z/\sqrt{2} & w/\sqrt{2} & -1/\sqrt{2} \\ -z/\sqrt{2} & w/\sqrt{2} & 1/\sqrt{2} \end{bmatrix}, \quad (1)$$

where $\chi_{23} = \pi/4$ and $x, y, \ldots$ are elements of a $SU(2)$ matrix. After setting $|x_1| = |w_1| = \cos\phi = \sqrt{2/3}$, $|y_1| = |z_1| = \sin\phi = \sqrt{1/3}$, we obtain the standard TBM matrix. The lower right 2 by 2 block in the first factor diagonalizes the unique $SU(1,1)$ element with real positive entries. The diagonalization generates the spectrum $m_0\left(e^{-\lambda}, e^{+\lambda}\right)$ of a $SU(1,1)$ flavor spin doublet, where $m_0$ is the doublet mass scale and $\lambda$ is the split parameter. For leptons at tree level only the $(\mu, \tau)$ doublet has non-zero $\lambda$. The upper left 2 by 2 block in the second factor is an $SU(2)$ matrix, which arises in the Cartan decomposition of the $SU(2,2)$ mass matrices. Further details can be found in [8 - 10].

Within the framework of the flavor spin theory the general form of the $TM_1$, was first proposed in [4] on purely theoretical grounds. It is obtained from (1) by multiplying $U_{TBM}$ from the right by the unique quantum correction term

$$U_{13} = \begin{bmatrix} 1 & 0 & 0 \\ 0 & \cos\chi_{13} & \sin\chi_{13}\, e^{-i\alpha_{CP}} \\ 0 & -\sin\chi_{13}\, e^{+i\alpha_{CP}} & \cos\chi_{13} \end{bmatrix}. \quad (2)$$

The result is

$$U_{ETBM} = \begin{bmatrix} \cos\chi_{12} & \cos\chi_{12}\sin\chi_{13} & \sin\chi_{12}\sin\chi_{13}\, e^{-i\alpha_{CP}} \\ -\sin\chi_{12}/\sqrt{2} & \left(\sin\chi_{13} + \cos\chi_{12}\cos\chi_{13}\, e^{-i\alpha_{CP}}\right)/\sqrt{2} & \left(-\cos\chi_{13} + \cos\chi_{12}\sin\chi_{13}\, e^{-i\alpha_{CP}}\right)/\sqrt{2} \\ -\sin\chi_{12}/\sqrt{2} & \left(-\sin\chi_{13} + \cos\chi_{12}\cos\chi_{13}\, e^{-i\alpha_{CP}}\right)/\sqrt{2} & \left(\cos\chi_{13} + \cos\chi_{12}\sin\chi_{13}\, e^{-i\alpha_{CP}}\right)/\sqrt{2} \end{bmatrix}. \quad (3)$$

The standard three-factor decomposition of $U_{PMNS}$ is

$$U_{PMNS}(\theta,\delta_{CP}) = \begin{bmatrix} c_{12}c_{13} & s_{12}c_{13} & s_{13}e^{-i\delta_{CP}} \\ -s_{12}c_{23} - c_{12}s_{13}s_{23}e^{i\delta_{CP}} & c_{12}c_{23} - s_{12}s_{13}s_{23}e^{i\delta_{CP}} & c_{13}s_{23} \\ s_{12}c_{23} - c_{12}s_{13}s_{23}e^{i\delta_{CP}} & -c_{12}c_{23} - s_{12}s_{13}s_{23}e^{i\delta_{CP}} & c_{13}c_{23} \end{bmatrix}. \quad (4)$$

Obviously, for absolute values of the matrix elements we have to agree and

$$|U_{ÜMNS}(\chi,\alpha_{CP})| = |U_{PMNS}(\theta,\delta_{CP})|. \quad (5)$$

Note, that because of re-phasing freedom we cannot assume $\alpha_{CP} = \delta_{CP}$ . As a result of unitarity, the matrix constraint (5) contains only four independent equations on the absolute values of the matrix elements. Therefore, we can express one set of four parameters in (5) in terms of the other. From (5) we obtain

$$c_{12}c_{13} = \cos\chi_{12}, \qquad s_{12}c_{13} = \cos\chi_{12}\sin\chi_{13}, \qquad s_{13} = \sin\chi_{12}\sin\chi_{13}, \quad (6)$$

from which we obtain

$$\sin^2\theta_{12} = \left(1 + \mathrm{ctg}^2\chi_{12}\cos^{-2}\chi_{13}\right)^{-1}, \qquad \sin^2\theta_{13} = \sin\chi_{12}\sin\chi_{13}. \quad (7)$$

These relations were first described in [2]. The two equations $|U_{\mu 1}(\chi)| = |U_{\mu 1}(\theta)|$ and $|U_{\tau 1}(\chi)| = |U_{\tau 1}(\theta)|$ can be used to obtain

$$\cos\delta_{CP} = \frac{1}{2}\left(\mathrm{ctg}\,\theta_{12}\sin\theta_{13} - \mathrm{tg}\,\theta_{12}\sin^{-1}\theta_{13}\right)\mathrm{ctg}(2\theta_{23}), \quad (8)$$

and from $|U_{\mu 3}(\chi)| = |U_{\mu 3}(\theta)|$, $|U_{\tau 3}(\chi)| = |U_{\tau 3}(\theta)|$ we obtain

$$\sin^2\theta_{23} = \frac{1}{2}\left(1 + \frac{\sin 2\chi_{13}\cos\chi_{12}}{\cos^2\chi_{13} + \cos^2\chi_{12}\sin^2\chi_{13}}\right)\cos\alpha_{CP}. \quad (9)$$

Equations (7-9) describe the transformation from the sets of parameters $(\chi,\alpha_{CP})$ to $(\theta,\delta_{CP})$.

Having derived the transformation, we can specify $TM_1$ matrix further using $|U_{e1}|^2 \approx 2/3$ and setting $\cos\chi_{12} = \sqrt{2/3}$ . In addition, we can use the apparently previously unreported relation that holds to the accuracy of $10^{-4}$

$$Q \equiv 4\left|U_{e2}\right|\left|U_{e3}\right| + \left|U_{e1}\right|^2 = 1. \quad (10)$$

Using its exact value in (10) we obtain that $\chi_{13} = \pi/12$ and hence

$$\sin\chi_{13} = \left[\frac{1}{2}\left(1 - \frac{\sqrt{3}}{2}\right)\right]^{1/2}, \qquad \cos\chi_{13} = \left[\frac{1}{2}\left(1 + \frac{\sqrt{3}}{2}\right)\right]^{1/2}. \quad (11)$$

Note that we set $\chi_{23} = \pi/4$, because that is the only possible value for a classical theory with fermions described by quantum differential forms. The end result of our specifications is

$$\sin^2\theta_{12} = \left(1 + 2\cos^{-2}(\pi/12)\right), \quad (12)$$

$$\sin^2\theta_{13} = \frac{1}{3}\sin^2(\pi/12), \quad (13)$$

$$\sin^2\theta_{23} = \frac{1}{2}\left(1 + \frac{1}{2}\frac{\sqrt{2/3}}{\cos^2(\pi/12) + (2/3)\sin^2(\pi/12)}\right)\cos\alpha_{CP}, \quad (14)$$

$$\cos\delta_{CP} = \frac{1}{2}\left(\mathrm{ctg}\,\theta_{12}\sin\theta_{13} - \mathrm{tg}\,\theta_{12}\sin^{-1}\theta_{13}\right)\mathrm{ctg}(2\theta_{23}). \quad (15)$$

From the four parameters in the standard representation unknown remain only $\theta_{23}$ and $\delta_{CP}$. Angle $\theta_{23}$ cannot be set in radicals, because a constraint that involves it is not known at the present time. Nevertheless, the relation (15) between $\theta_{23}$ and $\cos\delta_{CP}$ involves only radicals. Most importantly, it can be experimentally tested.

As we already mentioned above our goal is not to compute the experimental values of $U_{PMNS}$ per se, but to establish its approximation that can be considered as arising from the tree level Lagrangian. Let us now see how accurate the approximation is. For the values of $\theta_{12}, \theta_{13}$ we have [11, 12]

$$\sin^2\theta_{12}^{\ th} = 0.31810\ldots \qquad \sin^2\theta_{12}^{\ \exp} = 0.308^{+0.012}_{-0.011}, \quad (16)$$

$$\sin^2\theta_{13}^{\ th} = 0.022329\ldots \qquad \sin^2\theta_{13}^{\ \exp} = 0.02215^{+0.00056}_{-0.00058}, \quad (17)$$

$$Q^{th} = 1 \qquad Q^{\exp} = 1.0055 \pm 0.0005, \quad (18)$$

where for illustrative purposes we use the values for normal ordering with atmospheric and IC24 data. As the last test of our use of flavor spin formalism we test the relation (15).To do this we compute

$$W(\theta)=\frac{1}{2}\left(\operatorname{ctg}\theta_{12}\sin\theta_{13}-\operatorname{tg}\theta_{12}\sin^{-1}\theta_{13}\right)=-2.1760\ldots, \tag{19}$$

to obtain

$$\cos\delta_{CP}\approx-2.17601\operatorname{ctg}(2\theta_{23}). \tag{20}$$

We can now test this expression using the latest PDG data [12]. We use the format in Table 14.7 of [12 ], where normal and inverted ordering data sets are represented. We display only averages, the statistical deviations can be found in Table 14.7. The results can be seen in Table 1.

| | w/o SK-ATM & IC24 | w SK-ATM & IC24 | w SK-ATM & IC24 | w SK-ATM & IC24 |
|---|---|---|---|---|
| NO | $\Delta\chi^2=0.6$ | Best Fit Ordering | Best Fit Ordering | Best Fit Ordering |
| Param | bfp | bfp | bfp | bfp |
| $\theta_{23}$ | 48.5 | 43.3 | 43.4 | 48.4 |
| $\delta_{CP}{}^{ex}$ | 177 | 212 | 216 | 202 |
| $\cos\delta_{CP}{}^{ex}$ | -0.9986 | -0.8480 | -0.8090 | -0.9271 |
| $\cos\delta_{CP}{}^{th}$ | 0.2671 | -0.1292 | -0.1216 | 0.2594 |

| | w/o SK-ATM & IC24 | w SK-ATM & IC24 | w SK-ATM & IC24 | w SK-ATM & IC24 |
|---|---|---|---|---|
| IO | Best Fit Ordering | $\Delta\chi^2=6.1$ | $\Delta\chi^2=5.0$ | $\Delta\chi^2=7.7$ |
| Param | bfp | bfp | bfp | bfp |
| $\theta_{23}$ | 48.6 | 47.9 | 47.6 | 48.3 |
| $\delta_{CP}{}^{ex}$ | 285 | 274 | 274 | 270 |
| $\cos\delta_{CP}{}^{ex}$ | 0.2588 | 0.0697 | 0.0697 | 0 |
| $\cos\delta_{CP}{}^{th}$ | 0.2749 | 0.2210 | 0.1980 | 0.2517 |

**Table 1**: Comparison of the theoretical and experimental values of $\cos\delta_{CP}$ for NO/IO

We see from Table 1 that there is one data set for IO with $\delta_{CP}{}^{ex}=285$, where the experimental and theoretical values agree fairly well. Inverting (20) we can also compute $\sin^2\theta_{23}{}^{th}$ using $\delta_{CP}{}^{\exp}$ as the input. We now can compare the two values

$$\sin^2\theta_{23}{}^{th}=0.55920\ldots \qquad \sin^2\theta_{23}{}^{\exp}=0.562^{+0.012}_{-0.015}. \tag{21}$$

We see that the theoretical value for $\sin^2\theta_{23}$ lies within the one sigma error bound. Note, that for this data set $\delta_{CP}{}^{\exp}=(3/2)\pi+\varphi,\ \varphi=\pi/12$. The remaining measurements do not satisfy

(20) well.

We observe that taking $\delta_{CP} = (19/12)\pi$ results in a reasonable agreement with at least one data set from the experiments. It could be a coincidence, but we conjecture it is unlikely that this is the case. Our model selects one particular data set with a very good confidence. The increased accuracy of the experimental results coming from JUNO and their analysis [14 - 20] will hopefully confirm or deny our conjecture. If our conjecture is confirmed, we would achieve our goal of finding a good analytical approximation to $U_{PMNS}(\theta, \delta_{CP})$ expressible in radicals. This is a necessary step, before a realization of $TM_1$ can be constructed within the framework of flavor spin theory.

## 3. Summary

In summary, we have described a novel method of the known deconstruction of the lepton mixing matrix in its elemental factors. While reproducing the previously known relation between $\theta_{12}, \theta_{13}$, we derived a novel relation between $\delta_{CP}$ and the remaining three angles of the decomposition. Our derivation is done within the framework of the theory of quantum fermionic differential forms. We have shown that we can construct an approximation of full $U_{PMNS}$ in terms of radicals. Considered as the tree level approximation, such an approximation should be useful for having a fresh look at the details of the spontaneous symmetry breaking in the bi-spinor reformulation of the SM.

Acknowledgements: I would like to thank Vernon Barger, Shun Zhou, and Hsiang-nan Li for useful comments.